\documentstyle[preprint,aps,floats,epsfig]{revtex}

\def\appendix{\par
 \setcounter{section}{0}
 \setcounter{subsection}{0}
 \def\thesection{Appendix \Alph{section}}
 \def\theequation{\Alph{section}.\arabic{equation}}
 \setcounter{equation}{0}}
\begin{document}
\tightenlines
\renewcommand{\thefootnote}{\fnsymbol{footnote}}

\title{Borel--Pad\'e vs Borel--Weniger method: a QED and a QCD
example}

\author{G. Cveti\v c$^{a}$\footnote{
e-mail: cvetic@apctp.org}
and Ji-Young Yu$^{b}$\footnote{
e-mail: yu@dilbert.physik.uni-dortmund.de}
}

\address{ $^a$Asia Pacific Center for Theoretical Physics, 
Seoul 130-012, Korea}
\address{ $^b$Department of Physics, Dortmund University,
44221 Dortmund, Germany}

%

\maketitle

\renewcommand{\thefootnote}{\arabic{footnote}}

\begin{abstract}

Recently, Weniger (delta sequence) method has been
proposed by the authors of Ref.~\cite{JBWS} for resummation 
of truncated perturbation series in quantum field theories.
Those authors presented numerical evidence suggesting that
this method works better than Pad\'e approximants
when we resum a function with singularities
in the Borel plane but not on the positive axis.
We present here numerical evidence suggesting that 
in such cases the combined method of Borel--Pad\'e 
works better than its analog Borel--Weniger,
and that it may work better or comparably well
in some of the cases when there are singularities on the positive
axis in the Borel plane.\\
PACS number(s): 11.15.Bt, 10.10.Jj, 11.15.Tk, 11.80.Fv, 12.20.-m 

\end{abstract}

\section{Introduction}

In this letter we want to present some numerical 
results which allow us to compare the efficiency
of the Borel--Pad\'e method with that of the
Borel--Weniger method for resummation of
truncated perturbation series (TPS) in some 
physically significant scenarios.
The scenarios we are referring to are those
when the function, which we want to find through a resummation,
is known to have certain singularity structure
in the Borel plane. If there are singularities on the
positive axis of the Borel plane, then we implicitly
assume that in such cases we either know
the correct prescription for
integration in the Laplace--Borel integral, 
or we simply adhere to a certain adopted prescription.

(1) We will first illustrate the efficiency of the two methods
on the QED example of the
Euler--Heisenberg Lagrangian density, i.e.,
the one--loop fermion--induced effective action density
in a strong uniform electromagnetic field
\cite{Heisenberg:1936qt}--\cite{Itzykson:1980rh}.   
In this case, the solution is known, and its real part
can be written in the following form:
\begin{eqnarray}
{\rm Re} \delta {\tilde {\cal L}}({\tilde a};p) &=& 
{\rm Re} \int_0^{\infty}
d w \exp \left(- \frac{w}{\tilde a} \right)
\frac{(-1)}{w}
\left[ \frac{p \cos(w)}{\sin(w\!+\!{\rm i}{\epsilon}^{\prime})} 
\coth(p w) + \frac{1}{3} (1\!-\!p^2) - \frac{1}{w^2} \right] \ ,
\label{EH}
\end{eqnarray}
where we use notations
\begin{equation}
{\tilde a}\equiv\frac{g a}{m^2}, \ {\tilde b}\equiv\frac{g b}{m^2}, 
\qquad p \equiv \frac{b}{a} \equiv 
\frac{{\tilde b}}{{\tilde a}} \ , \qquad  
\delta {\tilde {\cal L}} \equiv 
\delta {\cal L}/ \left( \frac{m^4 {\tilde a}^2}{8 \pi^2} \right) \ .
\label{not}
\end{equation}
Here, $\delta {\cal L}$ is the actual Lagrangian
density induced by the one--loop fluctuations of the
fermions in the field;
$g$ is the field--to--fermion coupling parameter
(in QED it is the positron charge $e_0$); 
$m$ is the mass of the fermion (electron); 
$a$ and $b$ are Lorentz--invariant 
expressions characterizing the electric
and the magnetic fields ${\vec {\cal E}}$
and ${\vec {\cal B}}$, respectively
\begin{equation}
{a \choose b} =  \left[ \pm {\vec {\cal E}}^2 \mp {\vec {\cal B}}^2
+ \sqrt{ \left( {\vec {\cal E}}^2 - {\vec {\cal B}}^2 \right)^2
+ 4 \left( {\vec {\cal E}} \cdot {\vec {\cal B}} \right)^2 }
\right]^{1/2}/\sqrt{2} \ .
\label{ab}
\end{equation}
Expression (\ref{EH}) can be obtained, for example, directly by
integrating out the fermionic degreees of freedom in the path
integral expression of the full effective action, then employing
the proper--time integral representation for the difference
of logarithms, evaluating the traces in the integrand,
and subsequently performing Wick rotation
by $-\pi/4$ in the plane of the proper--time $s$: 
$a g s \mapsto -{\rm i} w \!+\!{\epsilon}^{\prime}$.
We refer to \cite{CY} for more details on the latter point. 
The perturbative expansion of the full solution (\ref{EH}), 
in powers of ${\tilde a}$, is
\begin{equation}
\delta {\tilde {\cal L}}^{\rm pert.}({\tilde a}; p) =
\left[
c_1(p) 1!\,{\tilde a}^2 + c_3(p) 3!\,{\tilde a}^4 + 
c_5(p) 5!\,{\tilde a}^6 + \cdots
\right] \ ,
\label{Lpert}
\end{equation}
with coefficients
\begin{eqnarray}
c_1(p) & = &  \frac{1}{45} \left[ 
(1\!-\!p^2)^2 + 7 p^2 \right] \ , \quad
c_3(p) =  \frac{1}{945} \left[ 
2 (1\!-\!p^2)^3 + 13 p^2 (1\!-\!p^2) \right] \ ,  \ {\rm etc.}
\label{cjs}
\end{eqnarray}
In the case of the pure magnetic field (p.m.f.), the corresponding
expressions are simpler
\begin{eqnarray}
\delta {\tilde {\cal L}}({\tilde b})_{{\tilde a}=0} 
\equiv \frac{ 8 \pi^2 \delta {\cal L}_{a=0} }{m^4 {\tilde b}^2}
=  \int_0^{\infty}
d w \exp \left( - \frac{w}{{\tilde b}} \right) \frac{(-1)}{w}
\left[ \frac{\coth(w)}{w} - \frac{1}{3} - \frac{1}{w^2} \right] \ ,
\label{EHpmf}
\end{eqnarray}
\begin{equation}
\delta {\tilde {\cal L}}^{\rm pert.}({\tilde b})_{{\tilde a}=0} =
\left[
{\tilde c}_1 1!\,{\tilde b}^2 + {\tilde c}_3 3!\,{\tilde b}^4 + 
\cdots \right] \ , \quad {\tilde c}_1\!=\!\frac{1}{45}, \
{\tilde c}_3\!=\!-\frac{2}{945}, \ldots
\label{Lpertpmf}
\end{equation}
We can now use (\ref{Lpert})--(\ref{cjs}), and (\ref{Lpertpmf}), 
as a laboratory for resummation methods, since the full (resummed)
solutions (\ref{EH}) and (\ref{EHpmf}) are known.
Since (\ref{EH}) and (\ref{EHpmf}) are Laplace--Borel integrals, 
it is natural to use these examples for testing
combined resummation techniques which involve 
Borel transformation. Borel transform $B_L$ of 
series (\ref{Lpert}) is
\begin{equation}
B_L(w;p) = c_1(p) w + c_3(p) w^3
+ c_5(p) w^5 + \cdots  \ ,
\label{BL}
\end{equation}
and analogously for (\ref{Lpertpmf}).
In Ref.~\cite{CY}, we used Borel--Pad\'e technique
for resummation, i.e., we applied various Pad\'e
approximants $[N/M]_B(w;p)$ to (\ref{BL})\footnote{
$[N/M]_B(w;p)$, being ratio of polynomials in
$w$ of powers $N$ and $M$, respectively \cite{Baker},
is based solely on the truncated perturbation
series (TPS) of (\ref{BL}) involving only terms with
$c_n$: $n \leq N\!+\!M$.}
and then employed the Laplace--Borel integral
to obtain the resummed value
\begin{equation}
BP^{\rm [N/M]} \left[ \delta {\tilde {\cal L}}^{\rm pert.}
\right]({\tilde a};p) =
\int_0^{\infty}
dw \exp \left( - \frac{w}{{\tilde a}} \right)
[N/M]_{\rm B}(w;p) \ .
\label{BPL}
\end{equation}
The integration over poles in (\ref{BPL}) was carried out
according to the Cauchy principal value prescription,
since the full solution (\ref{EH}) requires it.\footnote{
Various QCD and QED applications of the Borel--Pad\'e approach 
with the principal value prescription can be found in
\cite{Raczka}--\cite{Jentsch}. The novel method of
Ref.~\cite{Jentsch} is, in addition, well suited
for obtaining the imaginary part of $\delta {\cal L}$.
}

Recently, the authors of \cite{JBWS} proposed the
use of Weniger (delta sequence) transformations
as an alternative to the use of Pad\'e approximants,
for direct resummation of truncated perturbation series. 
For a truncated perturbation series (TPS) of the form
\begin{equation}
F_{[n+1]}(z) = \sum_0^{n+1} {\gamma}_j z^j
\label{TPS}
\end{equation}
it is defined as \cite{Weniger}
\begin{equation}
\delta_n^{(0)}(\zeta;\gamma_0,\ldots,\gamma_{n+1}) =
\frac{
\displaystyle{
\sum_j^n (-1)^j {n \choose j} 
\frac{(\zeta\!+\!j)_{n-1}}{(\zeta\!+\!n)_{n-1}}
\frac{z^{n-j} F_{[j]}(z)}{\gamma_{j+1}} 
}}
{
\displaystyle{
\sum_j^n (-1)^j {n \choose j} 
\frac{(\zeta\!+\!j)_{n-1}}{(\zeta\!+\!n)_{n-1}}
\frac{z^{n-j}}{\gamma_{j+1}} 
}} \ ,
\label{Wen}
\end{equation}
where $(\zeta\!+\!j)_{n-1} \equiv \Gamma(\zeta\!+\!j\!+\!n\!-\!1)/
\Gamma(\zeta\!+\!j)$ are the Pochhammer symbols and $\zeta\!=\!1$
is usually taken. The approximant (\ref{Wen}) is a ratio
of two polynomials in $z$ of power $n$ each, and when expanded back
in powers of $z$ it reproduces all the terms of $F_{[n+1]}$.

The authors \cite{JBWS} applied (\ref{Wen})
directly to the TPS's of 
$\delta {\tilde {\cal L}}^{\rm pert.}({\tilde a}; p)/{\tilde a}^2$
of (\ref{Lpert}), and when re-expanding the approximant
in powers of ${\tilde a}$ they were able to predict
the next coefficient in the series with a better precision
than the one provided by the corresponding diagonal
(or almost diagonal) Pad\'e approximant.
Further, in the case of the pure magnetic field 
they showed that the method (\ref{Wen}), when applied directly to the
TPS's in ${\tilde b}$ of the induced Lagrangian density,\footnote{
The approximants (\ref{Wen}) applied to the
TPS's of the series (\ref{Lpertpmf}) divided by ${\tilde b}^2$.}
gave better results of resummation than the corresponding
Pad\'e approximants.

We now combine the method (\ref{Wen}) with the
Borel transformation (\ref{Lpert}) $\mapsto$ (\ref{BL}), and compare the
results of resummation obtained in this way with the
results of the corresponding Borel--Pad\'e approximants
of Ref. \cite{CY}. Formula (\ref{Wen}) is applied
to the Borel transform (\ref{BL}) divided by $w$.
We identify $z\!\equiv\!w^2$ (we thank the authors of
\cite{JWS} for pointing out that this clarification
was missing in the original version of the preprint).
In the ensuing Borel--Weniger approximant, 
we integrate in the Laplace--Borel
integral over the poles of the integrand with the
Cauchy principal value prescription, just as 
in Borel--Pad\'e approximant (\ref{BPL}),
in accordance with the full known solution (\ref{EH}).

\begin{figure}
\noindent
\begin{minipage}[t]{.49\linewidth}
 \centering\epsfig{file=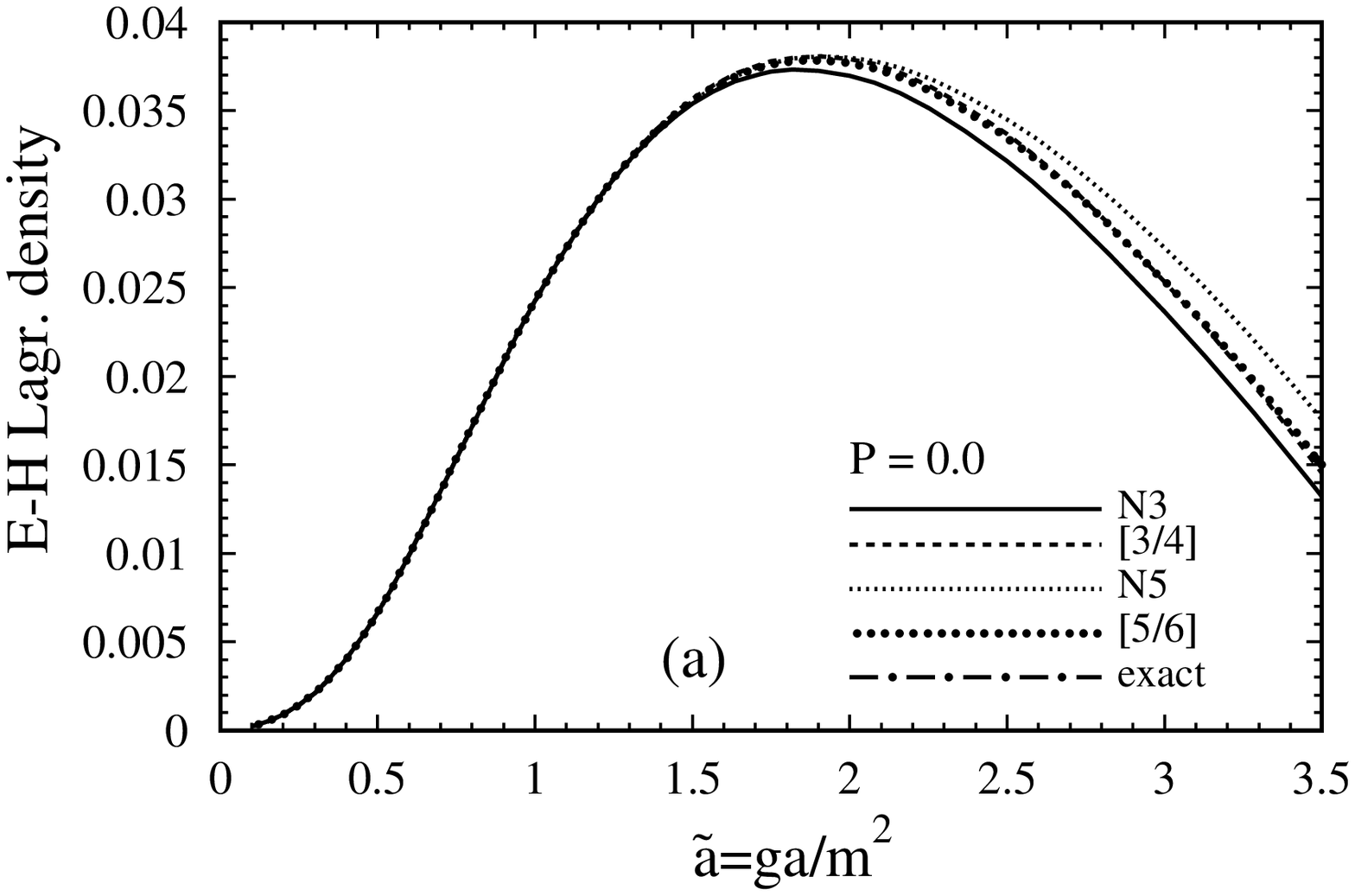,width=\linewidth}
 \centering\epsfig{file=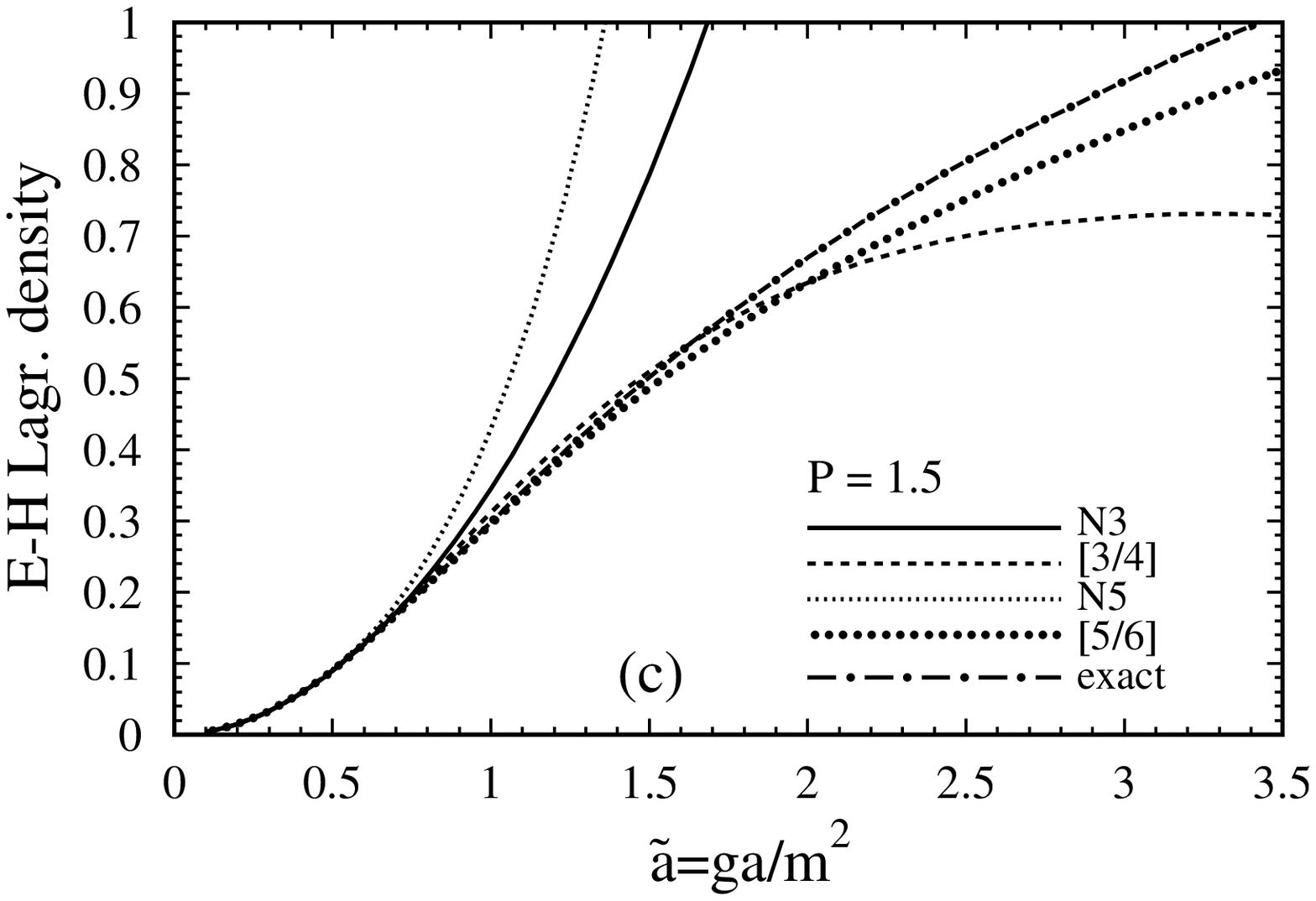,width=\linewidth}
\end{minipage}
\begin{minipage}[t]{.49\linewidth}
 \centering\epsfig{file=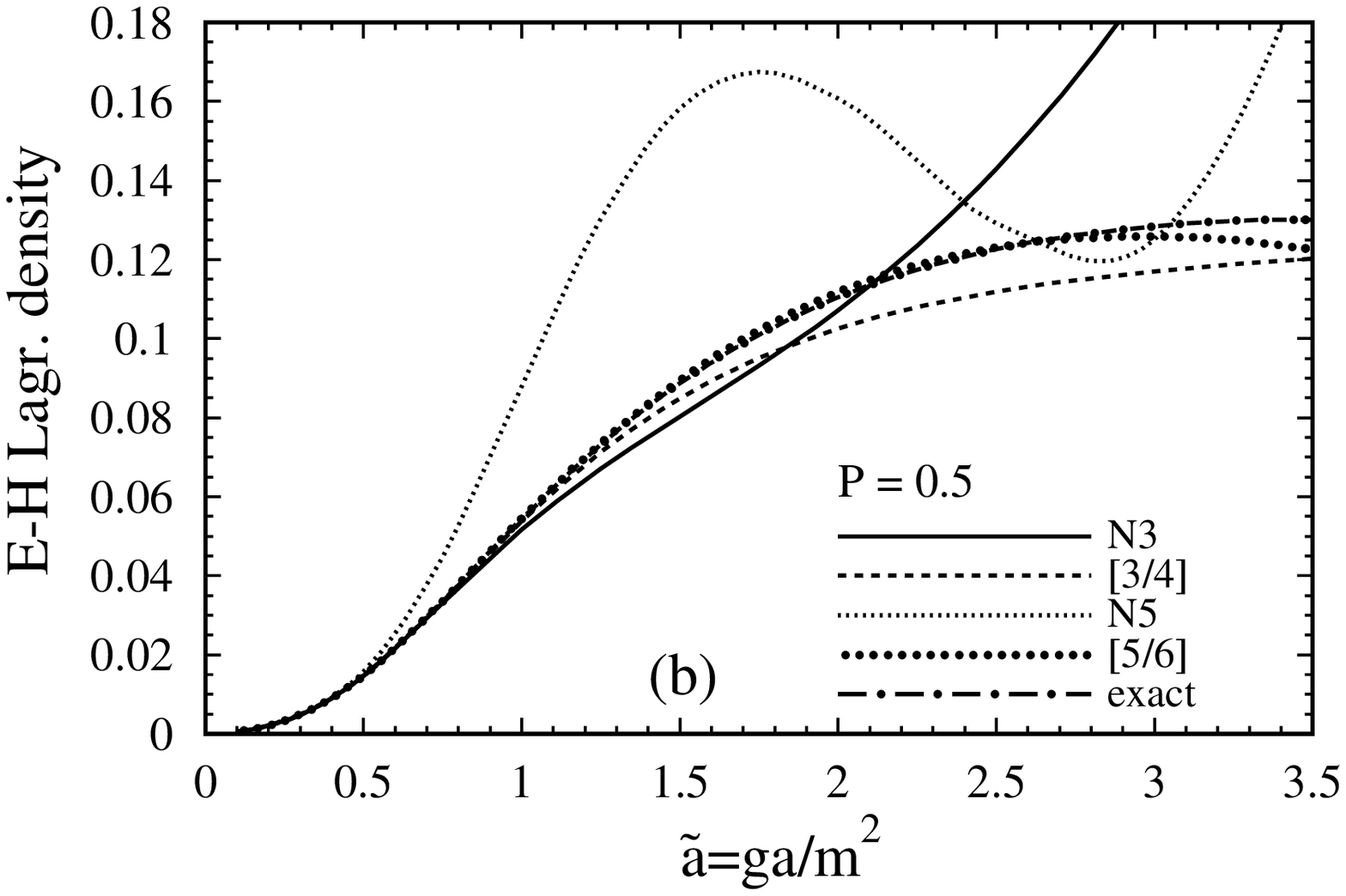,width=\linewidth}
 \centering\epsfig{file=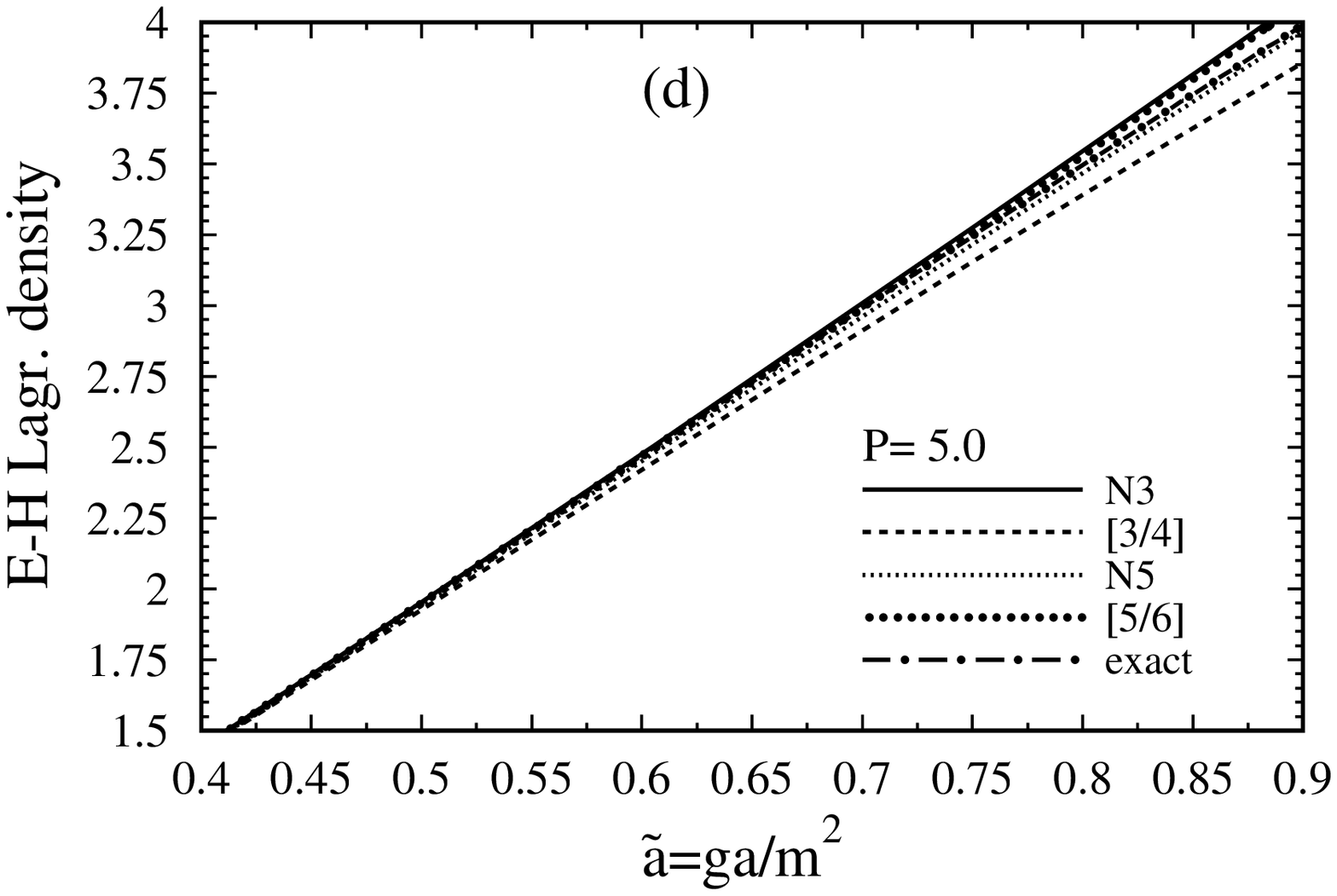,width=\linewidth}
\end{minipage}
\caption{\footnotesize
Borel--Pad\'e approximants ([3/4], [5/6])
and the corresponding
Borel--Weniger approximants (N3, N5) to the induced 
dispersive Lagrangian density (\ref{EH}), as functions
of ${\tilde a}$, for various values of
$p\!=\!{\tilde b}/{\tilde a}$: (a) $p\!=\!0.0$;
(b) $p\!=\!0.5$; (c) $p\!=\!1.5$; (d) $p\!=\!5.0$.
The numerically exact curves are included
for comparison.}
\label{Lel}
\end{figure}

\begin{figure}[htb]
\setlength{\unitlength}{1.cm}
\begin{center}
\epsfig{file=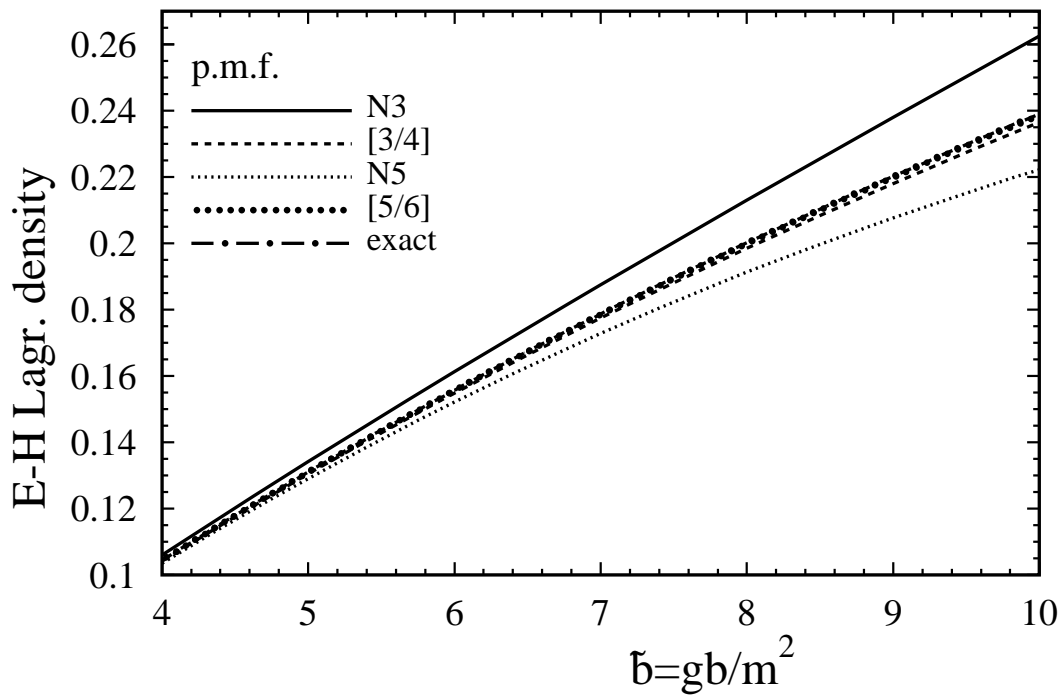,width=15.cm}
\end{center}
\vspace{-0.0cm}
\caption{\footnotesize Borel--Pad\'e approximants ([3/4], [5/6])
and Borel--Weniger approximants (N3, N5) to the induced 
dispersive Lagrangian density (\ref{EHpmf}), as functions
of ${\tilde b}$, for the pure magnetic field case
(${\tilde a}\!=\!0$). The numerically exact curve is included
for comparison.}
\label{Lmag}
\end{figure}

The results of these calculations are presented
in Figs.~\ref{Lel}(a)--(d), as functions of the electric
field strength parameter ${\tilde a}$, for various
values of $p\!\equiv\!{\tilde b}/{\tilde a} = 0., 0.5, 1.5, 5.0$.
In Fig.~\ref{Lmag} we present the analogous results for the
case of the pure magnetic field (p.m.f.), as function
of the magnetic field parameter ${\tilde b}$.
N3 and $[3/4]$ denote the Borel--Weniger 
and the Borel--Pad\'e resummations based on
the truncated Borel transform
(\ref{BL}) with the first four nonzero terms (i.e., three terms
beyond the leading order);
N5 and $[5/6]$ are based on the first six
terms in (\ref{BL}). Comparison with the exact
solutions, also present in the Figures, 
shows that Borel--Pad\'e is better than the corresponding 
Borel--Weniger, except in the case of $p\!=\!5.0$
(electric field combined with a much stronger magnetic field).
Fig.~\ref{Lmag} suggests that Borel--Pad\'e is better
than Borel--Weniger for resummation of functions whose
Borel transforms have singularities 
only outside the positive axis.
Further, comparison of Fig.~2 with the results of Table I
of Ref.~\cite{JBWS} suggests that 
Borel--Pad\'e and Borel--Weniger methods are much more
efficient than Weniger method in resumming series with
singularities in the Borel plane. Weniger method in 
the p.m.f.~case is better than Pad\'e method \cite{JBWS}.

\begin{figure}
\begin{minipage}[t]{.49\linewidth}
 \centering\epsfig{file=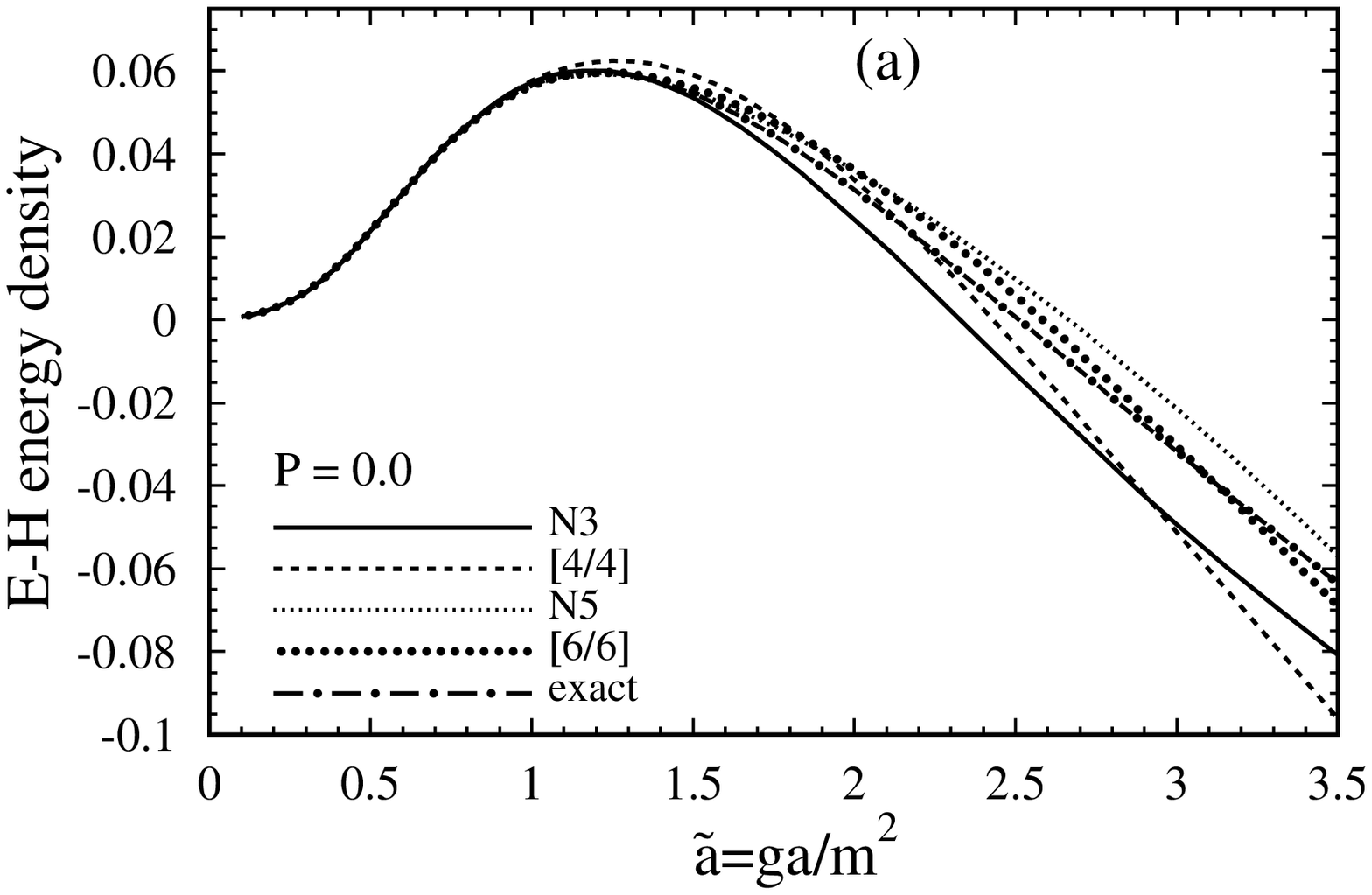,width=\linewidth}
 \centering\epsfig{file=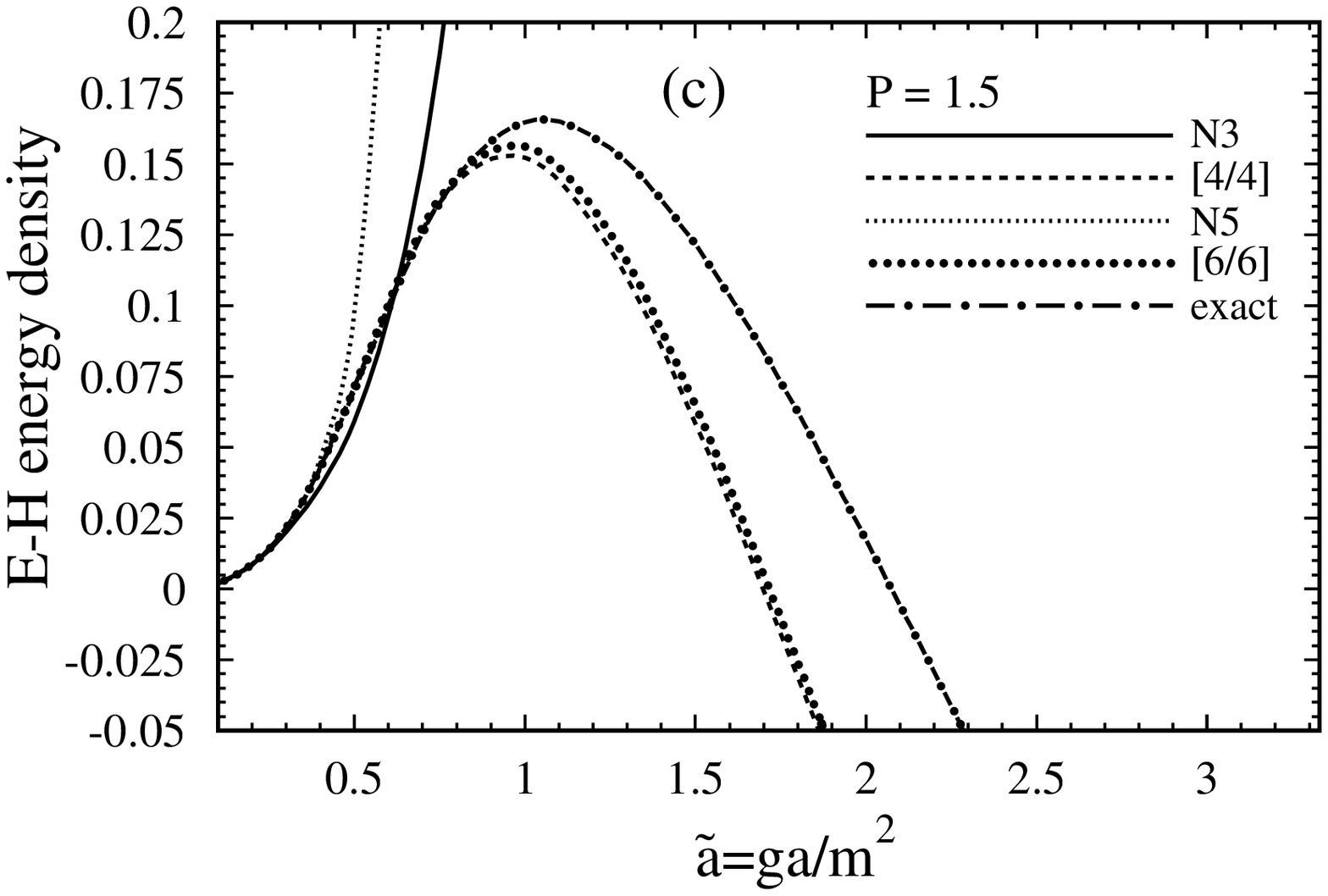,width=\linewidth}
\end{minipage}
\begin{minipage}[t]{.49\linewidth}
 \centering\epsfig{file=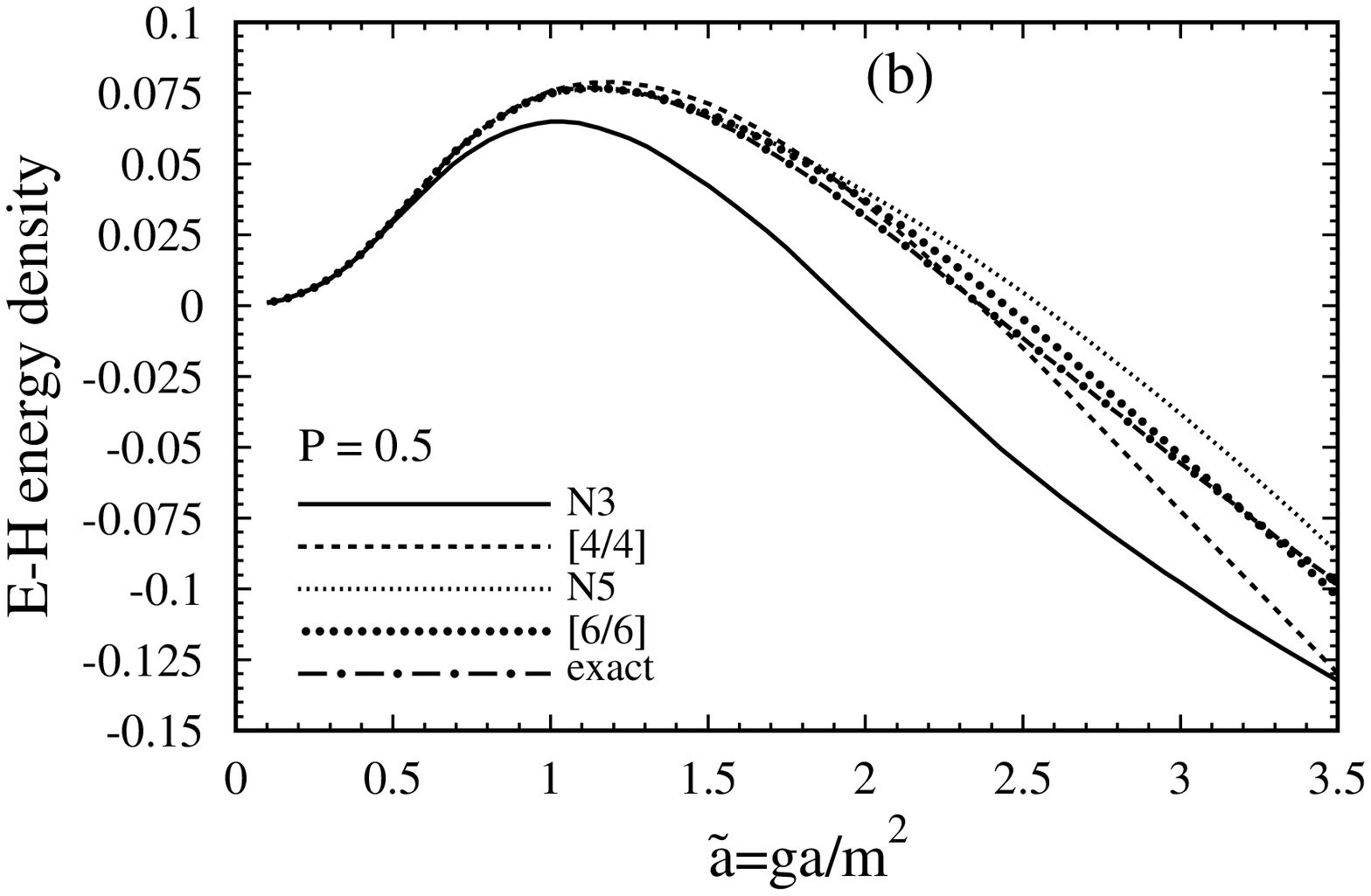,width=\linewidth}
 \centering\epsfig{file=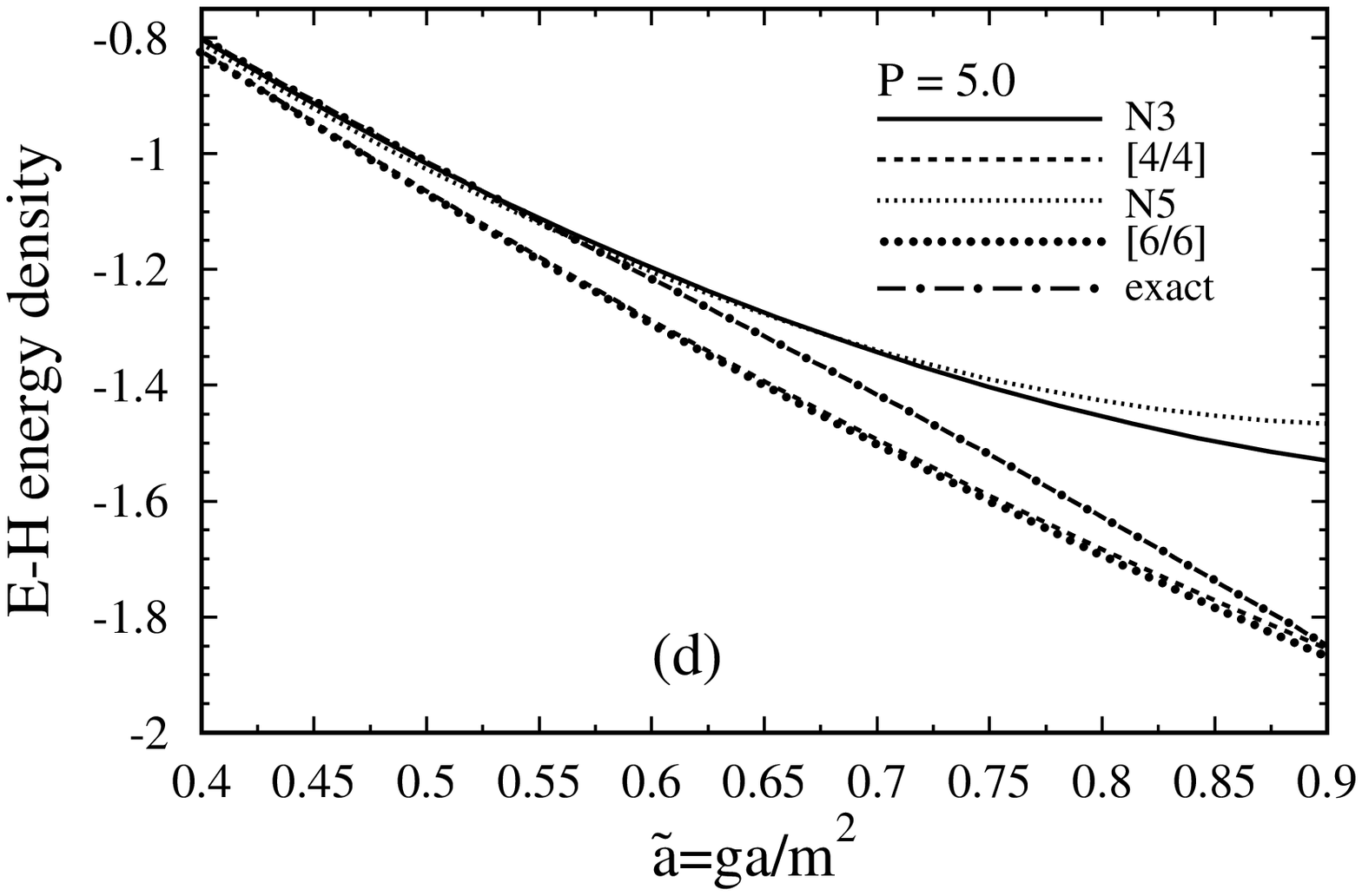,width=\linewidth}
\end{minipage}
\vspace{0.0cm}
\caption{\footnotesize
Modified Borel--Pad\'e ([4/4], [6/6])
and the corresponding modified Borel--Weniger (N3, N5)
approximants to the induced energy densities (\ref{U1}), 
as functions of ${\tilde a}$, at fixed
values of $p\!=\!{\tilde b}/{\tilde a}$:
(a) $p\!=\!0.0$; (b) $p\!=\!0.5$; (c) $p\!=\!1.5$; (d) $p\!=\!5.0$.}
\label{Uel}
\end{figure}

We can also do analogous calculations 
for the induced energy densities $\delta U$
\begin{eqnarray}
\delta {\cal U} &=&
a \frac{{\partial} {\rm Re} \delta {\cal L}}{{\partial} a}
{\Bigg |}_b 
- {\rm Re} \delta {\cal L} \ ,
\label{defdU}
\\
\delta {\tilde {\cal U}}({\tilde a}; p) &=&
{\rm Re} \int_0^{\infty}
d w \exp \left(- \frac{w}{\tilde a} \right)
\frac{(-1)}{w}
\left[ - \frac{ p w }{\sin^2(w\!+\!{\rm i}{\epsilon}^{\prime})} 
\coth(p w) + \frac{1}{3} (1\!+\!p^2) + \frac{1}{w^2} \right] \ ,
\label{U1}
\\
\delta {\tilde {\cal U}}^{\rm pert.} &=&
\left[
d_1(p) 1!\,{\tilde a}^2 + d_3(p) 3!\,{\tilde a}^4 + 
\cdots
\right] \ ,
\label{Upert}
\\
d_1(p) &=&  \frac{1}{45} \left[ 
3 + 5 p^2 - p^4 \right] \ , \quad
d_3(p) =  \frac{1}{945} \left[ 
10 + 21 p^2 - 7 p^4 + 2 p^6 \right] \ ,  \ {\rm etc.}
\label{djs}
\end{eqnarray}
where $\delta {\tilde {\cal U}}\!\equiv\!8 \pi^2 \delta {\cal U}/
(m^4 {\tilde a}^2)$. 
In that case, the simple Borel transform has a double--pole
structure on the positive real axis, and the Pad\'e and
Weniger approximants have trouble simulating such multiple
poles adequately. Therefore, we employ a slightly modified
Borel transform in the case of the induced energy densities
\begin{equation}
MB_U(w;p) = d_1(p) \frac{w^2}{2} + d_3(p) \frac{w^4}{4} + 
d_5(p) \frac{w^6}{6} + \cdots \ ,
\label{MB}
\end{equation}
which has no multiple--pole structure -- all the poles are simple.
The (modified) Laplace--Borel integral in this case is
\begin{equation}
\delta {\tilde {\cal U}}({\tilde a}; p) =
\frac{1}{{\tilde a}} \int_0^{\infty}\! dw 
\exp \left( - \frac{w}{{\tilde a}} \right) MB_U(w;p) \ ,
\label{MLBint}
\end{equation}
where again the Cauchy principal value has to be taken,
once $MB_U(w;p)$ is replaced in (\ref{MLBint}) by
its Pad\'e or Weniger approximants.
For details, we refer to Ref.~\cite{CY} where
Borel--Pad\'e was employed also for the induced energy densities. 
Weniger formula (\ref{Wen}) is now applied to the
modified Borel transform (\ref{MB}) divided by $w^2$.
The results are presented in Figs.~\ref{Uel}(a)--(d), as functions of
${\tilde a}$ at fixed $p\!=\!0., 0.5, 1.5, 5.0$, 
respectively.\footnote{
In the case of the pure magnetic field, the energy density
is the same as the Lagrangian density, except for the
sign change.} 
We present the solutions of
Borel--Weniger and Borel--Pad\'e based on the
first four (N3, [4/4]) and six (N5, [6/6]) nonzero terms
of the modified Borel transform of the energy density.
We see that for the induced energy density
the situation is less clear. In the cases
$p=0$, $0.5$ and $5.0$ the Borel--Pad\'e and Borel--Weniger
resummations are apparently of comparable
quality, while at $p=1.5$ the Borel--Pad\'e
appears to work better.

We can see these trends also if we compare the
perturbation coefficients predicted by these
two methods with the exact ones. These results are
written in Table \ref{tabl1} for the case of the
Lagrangian density (predicted $c_9$ and $c_{13}$)
and in Table \ref{tabl2} for the case of the energy density 
(predicted $d_9$ and $d_{13}$) . 
Predictions of Borel--Pad\'e and
Borel--Weniger are of comparable quality in the
cases of $p\!=\!0., 0.5, 5.0$ for energy density
and in the case of $p\!=\!5.0$ for Lagrangian density.
In other cases, predictions of Borel--Pad\'e are better.
\begin{table}[ht]
\par
\begin{center}
\begin{tabular}{l|| c  c  c  c }
approximant  & $p = 0.0$ and p.m.f & $p = 0.5$ & $p = 1.5$ & $p = 5.0$ \\
\hline \hline
N3    & $c_9=2.1666\!\cdot\!10^{-6}$ & $c_9 = 3.524\!\cdot\!10^{-6}$ &
$c_9 = 4.320\!\cdot\!10^{-4}$ & $c_9 = 596.91$ \\
\hline
[3/4] & $c_9=2.1637\!\cdot\!10^{-6}$ & $c_9 = 3.648\!\cdot\!10^{-6}$ &
$c_9 = 5.866\!\cdot\!10^{-4}$ & $c_9 = 595.28$ \\
\hline
exact & $c_9=2.1644\!\cdot\!10^{-6}$ & $c_9 = 3.711\!\cdot\!10^{-6}$ &
$c_9 = 6.166\!\cdot\!10^{-4}$ & $c_9 = 596.24$ \\
\hline
\hline
N5    & $c_{13}=2.2212\!\cdot\!10^{-8}$ & $c_{13}=3.725\!\cdot\!10^{-8}$ &
$c_{13} = 2.460\!\cdot\!10^{-5}$ & $c_{13} = 3823.65$ \\
\hline
[5/6] & $c_{13}=2.2215\!\cdot\!10^{-8}$ & $c_{13}=3.804\!\cdot\!10^{-8}$ &
$c_{13} = 3.157\!\cdot\!10^{-5}$ & $c_{13} = 3824.42$ \\
\hline
exact & $c_{13}=2.2215\!\cdot\!10^{-8}$ & $c_{13}=3.805\!\cdot\!10^{-8}$ &
$c_{13} = 3.161\!\cdot\!10^{-5}$ & $c_{13} = 3824.45$ \\
\end{tabular}
\end{center}
\caption {\footnotesize  Coefficients $c_9$ and $c_{13}$ of the
perturbation series for the induced Lagrangian density, as predicted
by various Borel--Weniger and Borel--Pad\'e approximants.
We include exact values for comparison.}
\label{tabl1}
\end{table}
\begin{table}[ht]
\par
\begin{center}
\begin{tabular}{l|| c  c  c  c }
approximant  & $p = 0.0$ & $p = 0.5$ & $p = 1.5$ & $p = 5.0$ \\
\hline \hline
N3    & $d_9=2.3752\!\cdot\!10^{-5}$ & $d_9 = 3.8124\!\cdot\!10^{-5}$ &
$d_9 = 3.312\!\cdot\!10^{-4}$ & $d_9 = -452.06$ \\
\hline
[4/4] & $d_9=2.3658\!\cdot\!10^{-5}$ & $d_9 = 3.7974\!\cdot\!10^{-5}$ &
$d_9 = 4.529\!\cdot\!10^{-6}$ & $c_9 = -458.32$ \\
\hline
exact & $d_9=2.3808\!\cdot\!10^{-5}$ & $d_9 = 3.8085\!\cdot\!10^{-5}$ &
$d_9 = 2.503\!\cdot\!10^{-5}$ & $c_9 = -464.01$ \\
\hline
\hline
N5    & $d_{13}=3.3319\!\cdot\!10^{-7}$ & $d_{13}=5.4289\!\cdot\!10^{-7}$ &
$c_{13} = 8.162\!\cdot\!10^{-5}$ & $c_{13} = -2977.3$ \\
\hline
[6/6] & $d_{13}=3.3309\!\cdot\!10^{-7}$ & $d_{13}=5.4291\!\cdot\!10^{-7}$ &
$c_{13} = -1.571\!\cdot\!10^{-6}$ & $c_{13} = -2991.7$ \\
\hline
exact & $d_{13}=3.3322\!\cdot\!10^{-7}$ & $d_{13}=5.4301\!\cdot\!10^{-7}$ &
$c_{13} = -2.537\!\cdot\!10^{-6}$ & $c_{13} = -2976.7$
\end{tabular}
\end{center}
\caption {\footnotesize  Coefficients $d_9$ and $d_{13}$ of the
perturbation series for the induced energy density, as predicted
by various Borel--Weniger and Borel--Pad\'e approximants.
For comparison, exact values are included as well.}
\label{tabl2}
\end{table}
In fact, in the case $p\!=\!5.0$ of the energy density,
the modified Borel--Weniger is slightly, but discernibly,
better than the modified Borel--Pad\'e.
Comparing predictions of Table \ref{tabl1} (for $p\!=\!0.0$)
with those of Tables II and III of Ref.~\cite{JBWS} 
suggests strongly that the discussed Borel--Pad\'e
and Borel--Weniger methods are better
than Weniger method in predicting the
coefficients $c_n$. Weniger method
is better than Pad\'e method in predicting
$c_n$'s \cite{JBWS}.

(2) The second example to compare the efficiency of the
Borel--Pad\'e and Borel--Weniger methods will be taken from
QCD, and it will have to do with the ``fixing'' of a
pole of a Borel transform rather than with a resummation.
We look at the Bjorken polarized sum rule (BjPSR),
which involves the isotriplet combination of the 
first moments over $x_{\rm Bj}$ of proton and
neutron polarized structure functions
\begin{equation}
\int_0^1 d x_{\rm Bj} \left[ g_1^{(p)} (x_{\rm Bj}; Q^2_{\rm ph})
- g_1^{(n)} (x_{\rm Bj}; Q^2_{\rm ph}) \right] =
 \frac{1}{6} |g_A| \left[ 1 - S(Q^2_{\rm ph}) \right] \ .
\label{BjPSR1}
\end{equation}
Here, $p^2\!=\!-Q^2_{\rm ph}$$<0$ is $\gamma^{\ast}$ 
momentum transfer. At $Q^2_{\rm ph}\!=\!3 {\rm GeV}^2$ 
where three quarks are assumed active ($n_f\!=\!3$), 
and if taking ${\overline {\rm MS}}$ scheme and
renormalization scale (RScl) $Q_0^2\!=\!Q^2_{\rm ph}$, 
we have the following TPS of the BjPSR observable
$S(Q^2_{\rm ph})$ available \cite{GLZN}--\cite{LV}:
\begin{eqnarray}
S_{[2]}(Q^2_{\rm ph}; Q^2_0 = Q^2_{\rm ph}; c_2^{\rm MS}, c_3^{\rm MS})
&=& a_0 ( 1 + 3.583 a_0 + 20.215 a_0^2) \ ,
\label{TPSBj}
\\
{\rm with:} \quad a_0 = 
a(\ln Q_0^2; c_2^{\rm MS}, c_3^{\rm MS}, \ldots) \ , \quad 
n_f&=&3 \ , \ c_2^{\rm MS} = 4.471, \ c_3^{\rm MS} = 20.99 \ .
\label{TPSBjnot}
\end{eqnarray}
Here we denoted by $a$ the strong coupling parameter
$a\!\equiv\!{\alpha}_s/\pi$. 

It is known from \cite{BK}--\cite{Ji} that the Borel transform
$B_S(z)$ of $S$ has the lowest positive pole at
$z_{\rm pole}\!=\!1/{\beta_0}\!=\!4/9$ (leading infrared
renormalon) and that this pole has a much
stronger residuum than the highest negative pole
at $z_{\rm pole}\!=\!-1/{\beta_0}$ (leading ultraviolet
renormalon). The question we raise here is:
How well can Pad\'e and Weniger approximants
to the Borel transform $B_S(z)$ determine
the next coefficient $r_3$ of the term $r_3 a_0^4$ 
in the TPS (\ref{TPSBj}),
via the requirement that $z_{\rm pole}\!=4/9$?
For that, we have to know well the actual $r_3$.
That term can be determined reasonably well on the basis of 
two approximants discussed in \cite{GC} -- the
effective charge approximant (ECH) ${\cal A}_S^{(\rm ECH)}(c_3)$
with $c_3\!\approx\!20.$, and another, also RScl-- and
scheme--independent approximant ${\cal A}_{S^2}^{1/2}(c_3)$
with $c_3\!\approx\!15.5$. These two approximants
give the correct location of the leading infrared renormalon pole,
and when we expand them back in powers of $a_0$ we obtain
$r_3\!\approx\!129.4$ and $r_3\!\approx\!130.8$,
respectively. Therefore, we can estimate with high
confidence the actual $r_3$: $r_3 = 130. \pm 1$.  

It is important to consider the RScl--
and scheme--invariant Borel transform
when we want to apply Pad\'e or Weniger approximants to it,
so that the predicted values of $r_3$ will be independent
of the RScl-- and scheme in which we work at the intermediate
stage. Such a Borel transform has been used
in \cite{BTinv}, and we use its variant ${\widetilde B}_S(z)$
as specified in \cite{GC} [cf. Eqs. (18)--(20) there]. 
Such a Borel transform reduces (up to a $z$--dependent 
nonsingular factor) to the usual Borel transform in the
approximation of the one--loop evolution.
The resulting power expansion
of ${\widetilde B}_S(z)$ up to $\sim$$z^3$ will depend on the
coefficient $r_3$
\begin{equation}
{\widetilde B}_S(z) = 1 + \frac{32}{81} (\gamma\!-\!1) y + 0.02078... y^2 +
\frac{8}{729}(-21.88... +\!\frac{1}{6}r_3) y^3 + {\cal O}(y^4) \ ,
\label{BTS}
\end{equation}
where $\gamma\!=\!0.577...$ is Euler constant, and
$y\!\equiv\!2 \beta_0 z$. If we apply $[2/1]$ and
$[1/2]$ Pad\'e approximants to the TPS (\ref{BTS})
and demand $z_{\rm pole}\!=\!1/{\beta}_0$ ($y_{\rm pole}\!=\!2$),
we obtain predictions $r_3\!=\!137.0$ and $r_3\!=\!128.0$,
respectively. The prediction of $[1/2]$ is significantly better,
and this could possibly be explained with the more involved
denominator structure of $[1/2]$ in comparison to $[2/1]$.
When applying to (\ref{BTS}) Weniger formula (\ref{Wen})
($\delta_2^{(0)}$ with $\zeta\!=\!1$), we obtain $r_3\!=\!135.3$. 
This is further away from the actual value of $130.\pm 1.$
than the prediction of $[1/2]$. In both $[1/2]$ and
$\delta_2^{(0)}$, the denominators are polynomials of
quadratic degree in $z$. 

To summarize this QCD example: We applied Pad\'e and Weniger
approximants to a (TPS of a) Borel transform of the Bjorken
polarized sum rule and demanded that the leading infrared renormalon pole
be reproduced correctly. Weniger approximant $\delta_2^{(0)}$
then apparently gives a somewhat worse prediction for the
next coefficient than the corresponding Pad\'e approximant $[1/2]$.

\vspace{0.7cm}

The work of G.C. was supported by the Korean
Science and Engineering Foundation (KOSEF).
The work of J.-Y.Y. was supported by the
German Federal Ministry of Science (BMBF).

\noindent

\end{document}